# Role of quantum statistics in multi-particle decay dynamics

Avi Marchewka* and Er'el Granot

Department of Electrical and Electronics Engineering, Ariel University Center of Samaria, Ariel, Israel

*email: avi.marchewka@gmail.com

## Abstract

We investigate the role of quantum statistics in the decay dynamics of a muti-particle state which is suddenly released from a confining potential. For an initial box-like trap ,the exact dynamics is presented for both the bosonic and fermionic cases. The time-evolution of two-particle correlations is characterized and some counterintuitive features are discussed. For instance, it is shown that although there are higher chance of finding the two bosons (as oppose to fermions) in the initial trap region, there is also a higher chance of finding them on two opposite sides of the trap (i.e., each particle propagates in opposite direction(than in the fermionic case *as if the two bosons repel each other*. The results are demonstrated by numerical simulations and analytically for the short-time approximation.

PACS indexes 03.65.-w,05.30-d

## Introduction

The scenario, in which particles are released from confining potentials, is common in recent theoretical and experimental [1-7] studies. In these experiments particles are initially localized in a spatial trap and then they are released and propagate freely. This well-known experimental setup can emulate other physical scenarios, such as the tunneling decay of alpha particles from a radioactive nucleus, the emission of a particle from a molecule and, of course, the slit experiment. Moreover, recent advances in atom cooling and trapping technologies make it possible to trap bosonic and fermionic gases[8-17] or even low atom-number (Fock) states [18] by optical means. These optical traps can be shut off almost instantaneous and therefore allow investigating the subsequent expansion dynamics.

For recent related theoretical works on *N* particles see Ref.[19] , for a couple of particles in a double well see Ref.[20] , in the presence of a nonlinear potential see Ref.[21] and for tunneling decay of Fermion see Ref.[22].

It is the object of this paper to investigate the *generic* escaping dynamics of pairs of particles. We investigate the two cases when the escaping particles are either bosons or fermions and compare between them.

## Generic Trap

To simulate the trapping process we define a spatial regime $|x| < a$, which can be regarded as a trap. Initially we assume that both particles are located in the trap, i.e., at $|x| < a$; hence, we can assume that the Schrödinger wave equation

$$i\frac{\partial \psi}{\partial t} = -\frac{\partial^2 \psi}{\partial x^2} + V(x,t)\psi \qquad (1)$$

(where the dimensionless units $\hbar = 1$ and $2m = 1$ were adopted) is initially governed by the potential

$$V(x, t<0) = \begin{cases} V_{in}(x) & |x| < a \\ \infty & |x| \geq a \end{cases} \qquad (2)$$

where $V_{in}(x)$ is an arbitrary potential.

However, for $t > 0$ the particles are released by reducing the trap's boundaries to

$$V(x, t \geq 0) = V_{in}(x) \qquad (3)$$

and the particles are free to escape from the trap to infinity.

The particular case where initially $V_{out} = \infty, V_{in} = 0$, and for $t > 0$ $V_{out} = V_{in} = 0$ is discussed in details below, but it should be stressed that the main conclusions are totally generic.

Now suppose that $\psi_1(x, t = 0)$ and $\psi_2(x, t = 0)$ are both eigenstates of the initial Schrödinger equation (i.e., with the initial potential Eqs.1 and 2), and $\psi_1(x, t)$ and $\psi_2(x, t)$ are their propagation in the presence of the new potential Eqs.1 and 3. Then it should be stressed that since $\psi_1(x, t = 0)$ and $\psi_2(x, t = 0)$ are orthogonal (since they are eigenstate of an Hermitian operator), and since the Schrödinger dynamics conserve orthogonality then $\psi_1(x, t)$ and $\psi_2(x, t)$ are orthogonal for any time $t$.

The symmetry of the joint wave function of two identical non-interacting particles depends on the nature of the particles, either fermions or bosons. In particular [2],

$$\varphi^F(x_1, x_2; t) \equiv 2^{-1/2}[\psi_1(x_1, t)\psi_2(x_2, t) - \psi_1(x_2, t)\psi_2(x_1, t)]$$

$$\varphi^B(x_1, x_2; t) \equiv 2^{-1/2}[\psi_1(x_1, t)\psi_2(x_2, t) + \psi_1(x_2, t)\psi_2(x_1, t)] \qquad (4)$$

where the superscripts "B" and "F" stand for Boson and fermions, respectively. The difference in the local probability density at points $x_1, x_2$ between the bosons and the fermions scenarios is

$$\delta(x_1, x_2; t) \equiv |\varphi^B(x_1, x_2; t)|^2 - |\varphi^F(x_1, x_2; t)|^2,$$

which can be written simply as

$$\delta(x_1, x_2; t) \equiv 2\Re\{\psi_1(x_1, t)\psi_2(x_2, t)\psi_1^*(x_2, t)\psi_2^*(x_1, t)\} \qquad (5)$$

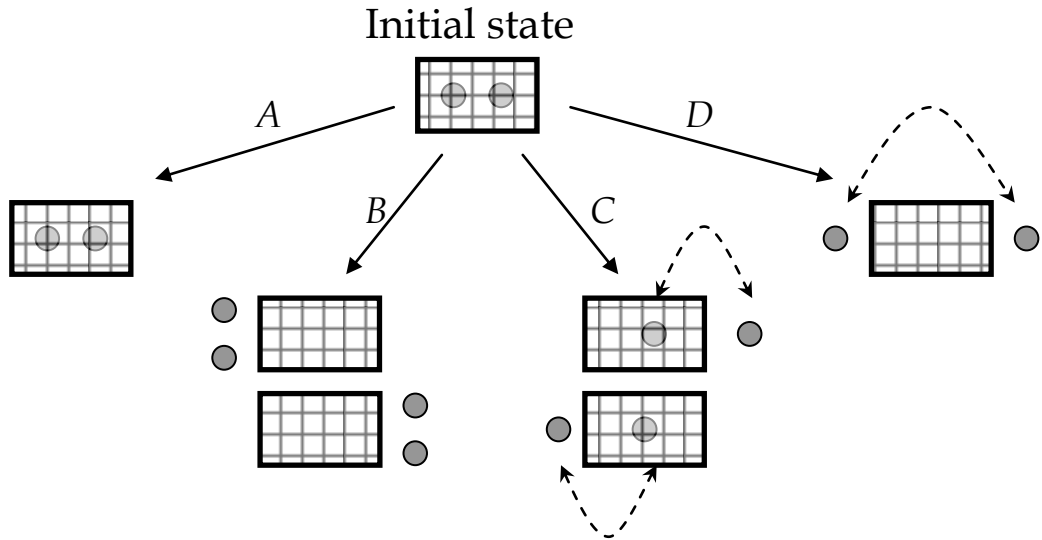

Figure 1: After releasing the particles from the trap – Four generic scenarios are possible: the two remains in the trap (A), the two escape form the trap to the same direction (B) only one escapes (C), the two escapes but to opposite directions

After releasing the two particles from the confining trap there are four possible scenarios:

A) The two particles remain in the trap (gray region).

B) The two particles escape form the trap, but they both escape to the same direction (black region).

C) Only one particle escapes from the trap – the other one remains inside (white region).

D) The two particles escapes from the trap, but to different direction (striped region).

Mathematically, the four scenarios are related to different integration regimes in the $x_1 - x_2$ map (see Fig.2)

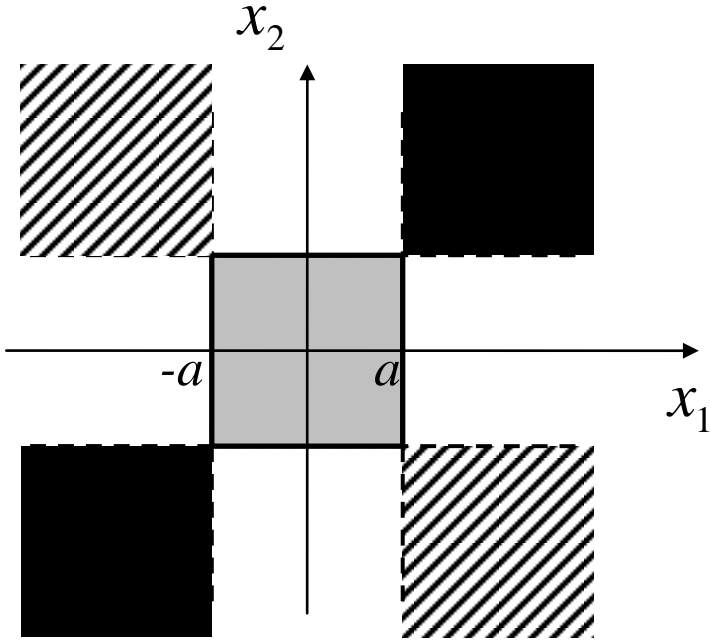

Figure 2: The four-region map in the $x_1 - x_2$ plane of the four scenarios (A-gray region, B-Black region, C- white region, and D- striped region)

Note that regions B (black) and D (striped) consist of two separated sections, whereas region C (white) consists of four separated ones. These sections are the manifestations of the problem's degeneracy.

For every one of these regions we define the probability difference

$$\Delta_S(t) \equiv \iint_S \delta(x_1, x_2; t) dx_1 dx_2$$

where $S = A, B, C,$ or $D$ and stands for the integration area in $x_1 - x_2$ plane.

$\Delta_S(t)$ can be regarded as a two-particle generalization of the escape probability . However, unlike Ref.19 they include three different types of escaping. In the following lines we are going to elaborate on the different scenarios, and calculate $\Delta(t)$ for each one of them.

***Scenario A*** – gray region (both particles remain in the trap):

$$\Delta_A(t) = 4\left|\int_{-a}^{a} dx \psi_2(x,t)\psi_1^*(x,t)\right|^2 \geq 0 \qquad (6)$$

That is – there is a higher chance of finding a bosons pairs in the trap region.

In particular, if the wave functions have different symmetries (one even and one odd), then $\Delta_A = 0$, which means that there is no preference between fermions or bosons to be remain in the trap.

***Scenario B*** – Black region (both particles escape to the same direction):

$$\Delta_B(t) = 4\left|\int_{a}^{\infty} dx \psi_2(x,t)\psi_1^*(x,t)\right|^2 + 4\left|\int_{-\infty}^{-a} dx \psi_2(x,t)\psi_1^*(x,t)\right|^2 \geq 0 \qquad (7)$$

Again, – there is a higher chance of finding bosons pairs escaping to the same direction.

In the case where the two wavefunctions have the same symmetry (either even or odd) then

$$\int_0^a dx \psi_2(x,t)\psi_1^*(x,t) = -\int_a^\infty dx \psi_2(x,t)\psi_1^*(x,t)$$

then

$$\Delta_B(t) = 8\left|\int_0^a dx \psi_2(x,t)\psi_1^*(x,t)\right|^2 = \frac{1}{2}\Delta_A(t) \geq 0 \qquad (8)$$

***Scenario D:*** As opposed to the previous scenarios (A and B), the sign of $\Delta_D$ depends on the symmetry of the problem. If the potential is symmetric, then its eigenfunctions are either symmetric or anti-symmetric (even or odd).

$$\Delta_D(t) = 2\Re\left\{\int_a^\infty dx_1 \psi_1(x_1,t)\psi_2^*(x_1,t)\int_{-\infty}^{-a} dx_2 \psi_2(x_2,t)\psi_1^*(x_2,t)\right\} +$$

$$2\Re\left\{\int_{-\infty}^{-a} dx_1 \psi_1(x_1,t)\psi_2^*(x_1,t)\int_a^\infty dx_2 \psi_2(x_2,t)\psi_1^*(x_2,t)\right\} \quad (9)$$

$$= \begin{cases} \Delta_B(t) & \text{same symmetry} \\ -\Delta_B(t) & \text{opposite symmetry} \end{cases}$$

That is, if the two wave functions have opposite symmetry (one even and one odd) then $\Delta_D(t) = -\Delta_B(t) \leq 0$, which means that Fermion would prefer to escape to opposite direction (as one would expect), however, if the two wavefunctions have the same symmetries (both are even or both are odd), then $\Delta_D(t) = \Delta_B(t) \geq 0$, *which means that bosons would prefer to escape to opposite direction.*

*Scenario C:* Since all wavefunctions are normalized then

$$\Delta_C(t) + \Delta_A(t) + \Delta_B(t) + \Delta_D(t) = 0 \quad (10)$$

Hence, if the two wave functions have opposite symmetry $\Delta_C(t) = \Delta_A(t) = 0$, which means that this scenario has no preference of Fermion over bosons or vice versa. On the other hand, when the wavefunctions have same symmetry then

$$\Delta_C(t) = -2\Delta_A(t) \leq 0 \quad (11)$$

which means that fermions prefer this scenario over bosons.

Table I: Summary of relations

| Symmetry\Region | A | B | C | D |
|---|---|---|---|---|
| **Same symmetry** | $\Delta_A(t) \geq 0$ | $\Delta_B = \Delta_A/2$ | $\Delta_C(t) = -2\Delta_A(t)$ | $\Delta_D(t) = \Delta_B(t)$ |
| **Opposite symmetry** | $\Delta_A(t) = 0$ | $\Delta_B(t) \geq 0$ | $\Delta_C(t) = \Delta_A(t) = 0$ | $\Delta_D(t) = -\Delta_B(t)$ |

Another important conclusion is that the total number of bosons in the trapping zone is equal to the total number of fermions there $N_B^{ToT} = N_F^{ToT}$, and therefore the total number of bosons outside the trapping zone is equal to the total number of fermions there too.

Therefore, one cannot distinguish between fermions or bosons by simply measuring the total *number* of particles in the trap, while such a distinction can be done by measuring pairs of particles. That means that a linear detector, which is placed in the trapping zone cannot distinguish between the different populations, while a nonlinear one (which is activated only by a couple of particles) can. Nevertheless, the specific scenario can be detectable by correlation between couple of detectors.

The scenarios A and the first part of C are the only ones, which are not surprising. It is not surprising that bosons have higher chances to be found in any region in space, and similarly, it is not surprising that fermions would prefer to be found separated (one will remain in the trap and the other will escape form it). However, it is very surprising that there are more chances of finding a couple of bosons outside the trap than to find a couple of fermions. In particular, if the wavefunctions have the same symmetry then there are higher chances of finding two repelling bosons (*escaping to opposite directions*) than repelling fermions (scenario D).

This shows that the statement on fermions repelling each other more than bosons do is a superficial statement. The repelling-attracting forces between them are more intricate.

Moreover, it can easily be shown that the average number of bosons (pairs or singles), which remain in the trap is equal to that of Fermion, regardless of the wavefunctions symmetry.

An interesting property of $\delta(x_1,x_2;t)$ when the two wavefunctions have opposite symmetry is that it flips sign under 90 degree rotation in the $x_1 - x_2$ plane, which means that under 90 degree rotation in the $x_1 - x_2$ plane $|\varphi^B(x_1,x_2;t)|^2$ becomes $|\varphi^F(x_1,x_2;t)|^2$ and vice versa. This is consistent with the fact that $\Delta_B = -\Delta_D$ and $\Delta_A = \Delta_C = 0$ in the opposite symmetry case.

## Escape from an infinite well

To illustrate the effect more clearly we consider the case where two pairs of particles, fermions and bosons, are initially in an infinite trap. Then, suddenly the trap is open and the particles propagates freely in space.

From the description of the previous section, four scenarios are possible after releasing the couple from the trap, A–D cases. To evaluate the probabilities

$$P^F(x_1,x_2;t) = |\psi^F(x_1,x_2;t)|^2, \quad P^B(x_1,x_2;t) = |\psi^B(x_1,x_2;t)|^2 \qquad (12)$$

of each one of the scenarios we measure the probability density in every region of the 2D map. First, the exact solution is presented, then a short-time approximation is introduced, and finally a full 2D map is plotted for different times.

We choose two symmetric eigenstates of an infinitely deep well as the initial condition. That is, we choose the Schrödinger equation (Eq.1) with the potential

$$V(x,t) = \begin{cases} \begin{cases} \infty & |x| > a \\ 0 & else \end{cases} & t < 0 \\ 0 & t > 0 \end{cases} \qquad (13)$$

For the initial wavefunctions (which are also eigenstates of the potential) we choose the symmetric eigenstates

$$\psi_j(x, t=0) = a^{-1/2} \cos(k_j x)[\Theta(-x+a) - \Theta(-x-a)] \qquad (14)$$

for $j = 1$ or $2$, and $k_1 = \pi/2a$, $k_2 = 3k_1$, which means that the first state is the ground state, and the second one is the third state (both are symmetric).

Then, for any $t > 0$ the dynamics resembles two Moshinsky's shutters [24-25],

$$\psi_j(x, t>0) = \frac{1}{2} a^{-1/2} \{\exp(ik_j a) M(x-a, k_j, t) - \exp(-ik_j a) M(x+a, k_j, t) + \\ \exp(-ik_j a) M(x-a, -k_j, t) - \exp(ik_j a) M(x+a, -k_j, t)\} \qquad (15)$$

where $M(x, k; t) \equiv \frac{1}{2} \exp(ikx - ik^2 t) \mathbf{erfc}\left(\frac{x - 2kt}{2\sqrt{it}}\right)$ is the Moshinsky function[25,26].

Since we know the dynamics of the two wavefunctions, we can easily find the difference in the density populations at any time (this will be illustrated graphically in the next section), however, at short times the difference is particularly large. In this section we will focus only on the scenario where both particles escape from the trap (scenario B). At short time (i.e., immediately after the particles are released from the trap) $x^2/t \to \infty$

$$\psi_j(x > a, x < -a, t > 0) \sim 2a^{-1/2} \sqrt{\frac{-i}{\pi}} \frac{t^{3/2}}{x^2} k_j \left[G_1(x) + \frac{k_j^2 t^2}{x^2} G_2(x)\right] \qquad (16)$$

where

$$G_1(x) \equiv x^2 \left\{ \frac{\exp\left[i\frac{(x+a)^2}{4t}\right]}{(x+a)^2} - \frac{\exp\left[i\frac{(x-a)^2}{4t}\right]}{(x-a)^2} \right\} \sim 2i\exp\left[i\frac{x^2+a^2}{4t}\right]\sin\left(\frac{xa}{2t}\right), \qquad (17)$$

and

$$G_2(x) = 4x^4 \left\{ \frac{\exp\left[i\frac{(x+a)^2}{4t}\right]}{(x+a)^4} - \frac{\exp\left[i\frac{(x-a)^2}{4t}\right]}{(x-a)^4} \right\} \sim 8i\exp\left[i\frac{x^2+a^2}{4t}\right]\sin\left(\frac{xa}{2t}\right). \qquad (18)$$

where the approximations on the right-hand side of Eqs.[17,18] were derived for $x \gg a$.

Hence, at short time the Fermionic wavefunction reads (for $|x_1|^2, |x_2|^2 \gg a \gg \sqrt{t}$)

$$\varphi^F(x_1, x_2; t) \equiv 2^{-1/2}[\psi_1(x_1)\psi_2(x_2) - \psi_1(x_2)\psi_2(x_1)] \sim$$
$$-i\frac{96}{\pi a}\frac{2^{-1/2}t^5}{(x_1 x_2)^2} k_1^4 \left( \frac{-1}{x_1^2} G_1(x_2)G_2(x_1) + \frac{1}{x_2^2} G_1(x_1)G_2(x_2) \right) \qquad (19)$$

which for $x \gg a$ can be simplified as

$$\psi^F(x_1, x_2; t) \sim i\frac{1536}{\pi a}\frac{2^{-1/2}t^5}{(x_1 x_2)^2} k_1^4 \exp\left[i\frac{x_1^2 + x_2^2 + 2a^2}{2t}\right]\sin\left(\frac{x_1 a}{2t}\right)\sin\left(\frac{x_2 a}{2t}\right)\left(\frac{1}{x_2^2} - \frac{1}{x_1^2}\right) \qquad (20)$$

while the Bosonic wavefunction reads (for $|x_1|^2, |x_2|^2 \gg a \gg \sqrt{t}$)

$$\varphi^B(x_1, x_2; t) \equiv 2^{-1/2}[\psi_1(x_1)\psi_2(x_2) + \psi_1(x_2)\psi_2(x_1)] \sim -i\frac{2a}{\pi a}\frac{2^{-1/2}t^3}{(x_1 x_2)^2} k_1^2 G_1(x_1)G_1(x_2) \qquad (21)$$

which for $x \gg a$ can be simplified to

$$\varphi^B(x_1, x_2; t) \sim i\frac{384}{\pi a}\frac{2^{-1/2}t^3}{(x_1 x_2)^2} k_1^2 \exp\left[i\frac{x_1^2 + x_2^2 + 2a^2}{4t}\right]\sin\left(\frac{x_1 a}{2t}\right)\sin\left(\frac{x_2 a}{2t}\right). \qquad (22)$$

We therefore see, that for short times, i.e., $t \ll x^2$, not only that the probability density of finding the two bosons outside the trap is larger than the probability density of finding the two fermions, but the two probability densities increase differently. The bosons probability density outside the trap is proportional to $t^6$, while the fermions one is proportional only to $t^{10}$.

Moreover, the probability density due to Eq.22 for bosons escape is proportional to the product of the particles' energies, while in the case of fermions it is proportional to the *square* of the same product. In this respect it should be noted that it was shown in [22] that the energy dispersion of the bosonic state is actually larger than in its fermionic counterpart.

Hence, at short time $|\varphi^B(x_1, x_2; t)|^2 \gg |\varphi^F(x_1, x_2; t)|^2$, which means that in regions B and D, $\delta(x_1, x_2; t) > 0$ and therefore $\Delta_B = \Delta_D > 0$, according to the generic theory.

The ratio between the two probability densities is

$$\frac{|\psi^F(x_1, x_2; t)|^2}{|\psi^B(x_1, x_2; t)|^2} \sim (2tk_1)^4 \left(\frac{1}{x_2^2} - \frac{1}{x_1^2}\right)^2 \ll 1 \tag{23}$$

which emphasizes that at short times the difference in population is enormous. As a pair bosons will prefer to escape from the trap – *even if they escape to opposite directions.*

Note, however, that in the case where the eigenstates have opposite symmetry there will be no difference between the two populations, as we explained above, and will be demonstrated in the next section.

It should be stressed that qualitatively, the conclusions of this section are generic. Indeed, the specific $t^4$ behavior of (23) is a consequence of the initial singularity (discontinuity of the wavefunction derivative), however, the general phenomenon is, as can be understood from the previous sections, generic.

It has been shown elsewhere[24], that the temporal dependence is affected only by the order of the singularity, hence, for any trap with infinite barriers (and therefore discontinuous wavefunction derivatives) the same $t^4$ behavior should appear. Moreover, it was demonstrated in the same place that even if the wavefunction has a smooth but sharp change in its derivatives' values then the same conclusions can be applied provided the measurements is not taken too far from the boundary (in Ref.[24] it was shown explicitly to discontinuous functions, but the analogy for discontinuous derivatives is straightforward). Obviously, this conduct disappears when the potential is smooth.

Since these results have only statistical meaning, the experiments should be repeated and the different scenarios frequencies should be compared.

It should be emphasized that our results are not restricted to the short-time approximation discussed in this section, but hold as well for arbitrary times as we shall demonstrated numerically in the following section.

## Numerical Simulations

The differences between the two probability densities can be illustrated graphically. In Fig.3 we plot graphically the Fermionic $P^F(x_1, x_2; t) = |\psi^F(x_1, x_2; t)|^2$ as well as the Bosonic $P^B(x_1, x_2; t) = |\psi^B(x_1, x_2; t)|^2$ joint probabilities initially (at t=0) and at t=0.03 for the wavefunctions, which have the same symmetry (Eq.14). The rectangle stands for the size of the initial trap ($|x_1| < 1/2, |x_2| < 1/2$)

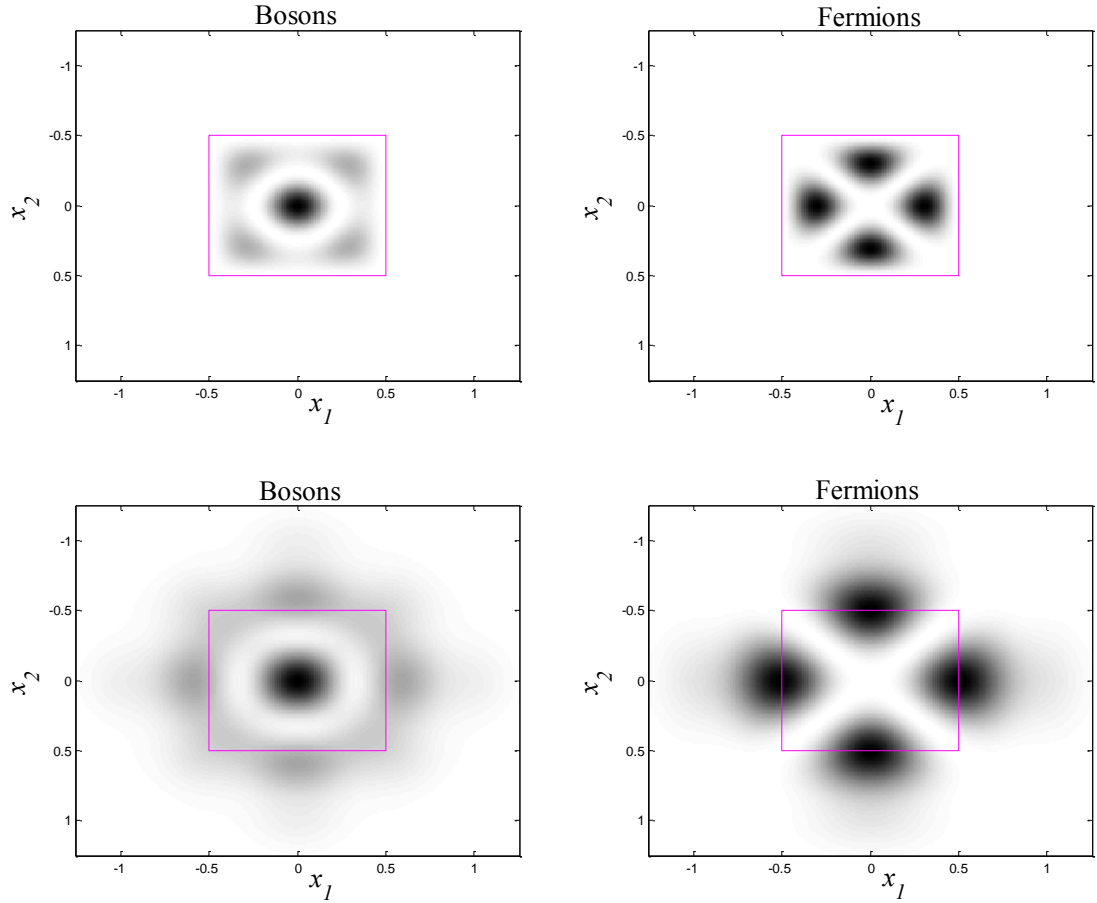

Figure 3 (color online): Gray levels presentation of $P^F(x_1, x_2; t=0)$ (upper right), $P^B(x_1, x_2; t=0)$ (upper left), $P^F(x_1, x_2; t=0.03)$ (lower right) $P^B(x_1, x_2; t=0.03)$ (lower left) for wavefunctions with the same symmetry (Eq. 14).

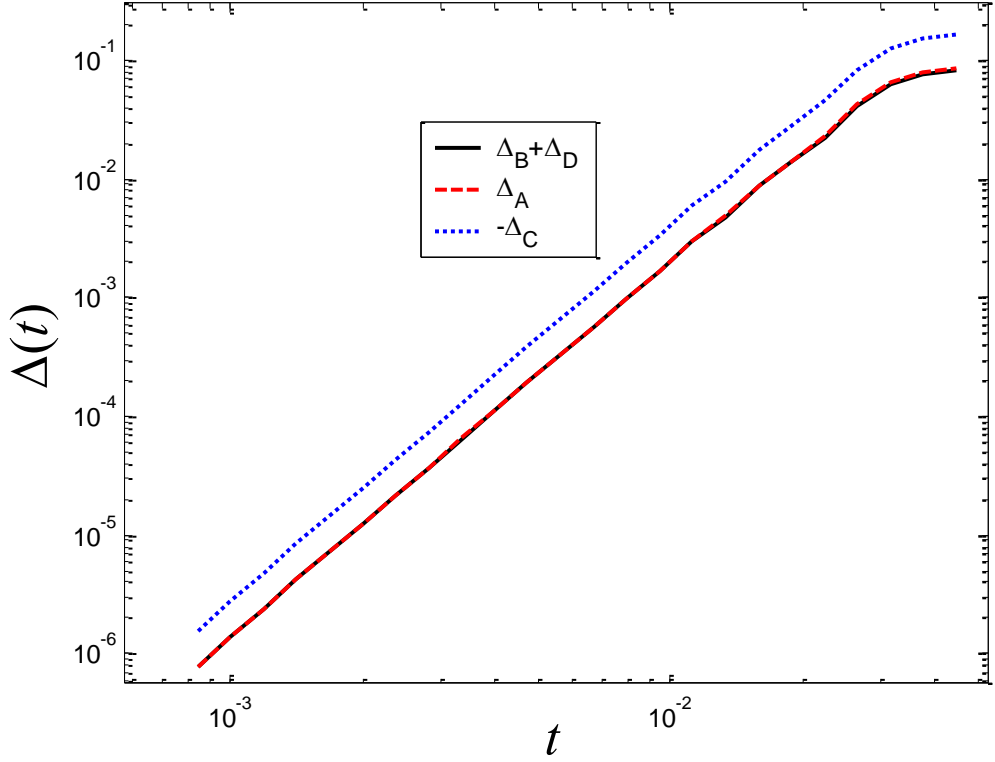

Figure 4 (color online): Temporal plots of $\Delta_B(t)+\Delta_D(t)$, $\Delta_A(t)$ and $-\Delta_C(t)$ (solid curve, dashed curve, and dotted curve respectively) for wavefunctions with the same symmetry. Cleary, $-\Delta_C(t) = 2\Delta_A(t)$

In Fig.4 we can see numerically that when the two wavefunctions are eignestates of the same potential and have the same symmetry (i.e. orthogonal) then $\Delta_A = \Delta_B + \Delta_D > 0$.

On the other hand, to illustrate that initial wavefunctions that have opposite symmetries behave totally differently we choose the following wavefunctions:

$$\psi_1(x, t=0) = a^{-1/2} \cos(k_1 x)[\Theta(-x+a) - \Theta(-x-a)] \quad (24)$$

with $k_1 = 3\pi/2a$, and

$$\psi_2(x, t=0) = a^{-1/2} \sin(k_2 x)[\Theta(-x+a) - \Theta(-x-a)] \quad (25)$$

with $k_2 = \pi/a$.

In Fig.5 we plot graphically the Fermionic $P^F(x_1,x_2;t)=|\psi^F(x_1,x_2;t)|^2$ as well as the Bosonic $P^B(x_1,x_2;t)=|\psi^B(x_1,x_2;t)|^2$ joint probabilities initially (at $t=0$) and at t=0.03 for these wavefunctions (Eqs.24 and 25), which have opposite symmetries.

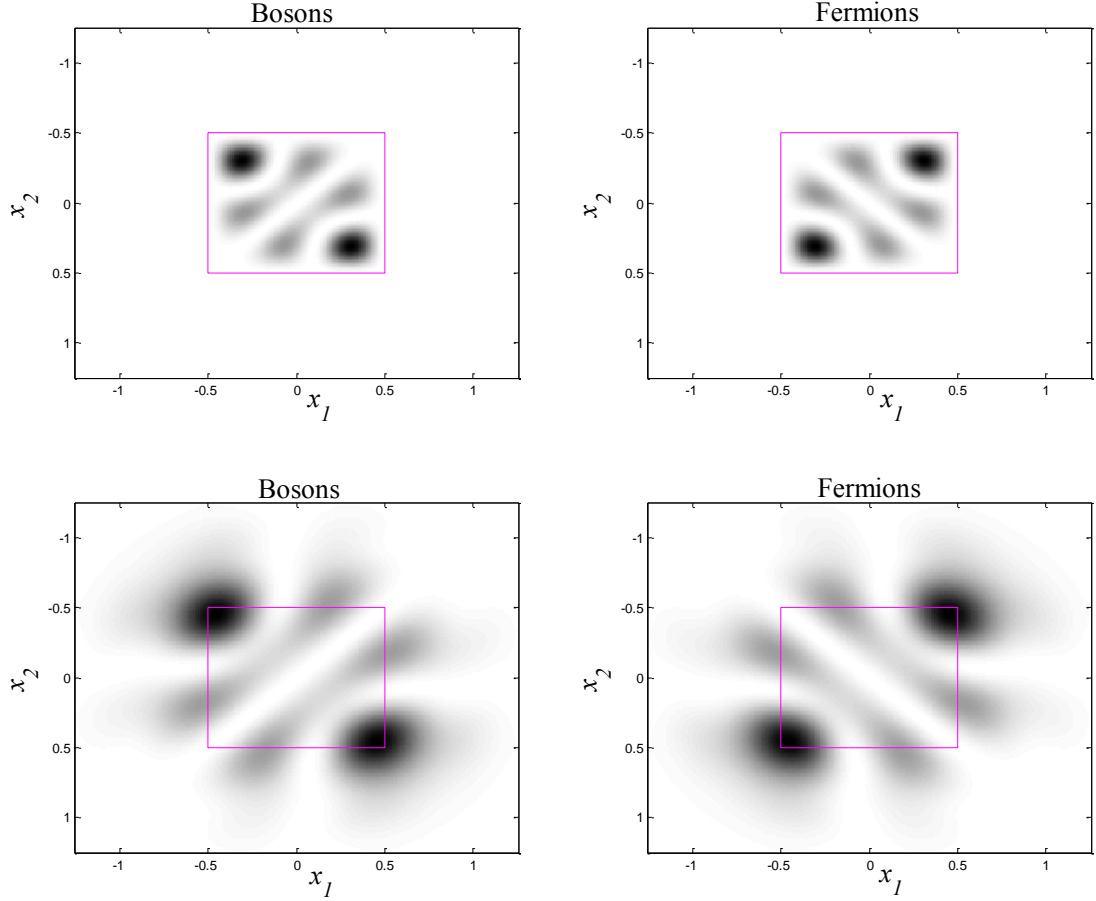

Figure 5: Gray levels presentation of $P^F(x_1,x_2;t=0)$ (upper right), $P^B(x_1,x_2;t=0)$ (upper left), $P^F(x_1,x_2;t=0.03)$ (lower right) $P^B(x_1,x_2;t=0.03)$ (lower left) for wavefunctions with different symmetry (Eqs.24 and 25).

The 90 degree rotation of the $x_1-x_2$ map is clearly shown, from which it is clear that $\Delta_A=\Delta_C=0$ and $\Delta_D=-\Delta_B$. As a result, all the interesting differences between the two populations, which were found in the same symmetry case, are lost in the opposite symmetry case.

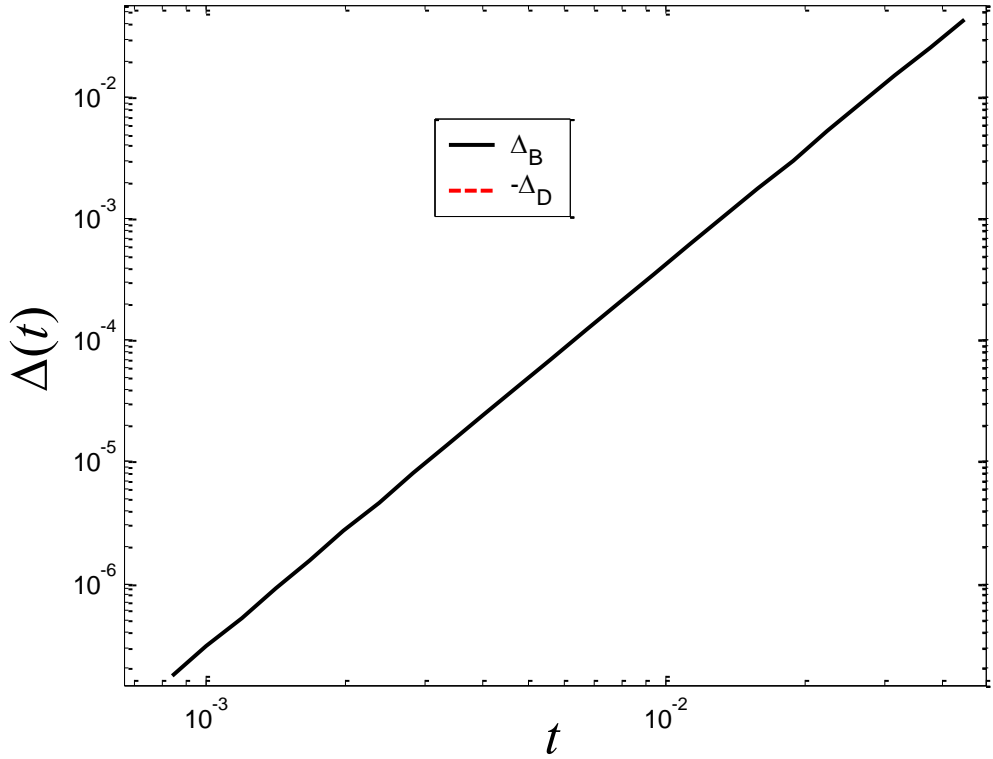

Figure 6 (color online): Temporal plots of $\Delta_B(t)$, and $-\Delta_D(t)$ (the two curves merge into a single line) for wavefunctions with different symmetry. Note that $\Delta_A(t) = \Delta_C(t) = 0$.

In Fig.7 the difference between the populations is presented for $t = 0.03$, i.e., $\delta(x_1, x_2; t = 0.03)$. In the case where the wavefunctions have the same symmetry (the left one) it is clear that there is no difference between the B and the D scenarios in both there is a preference for bosons. In this case, fermions would prefer scenario C, while bosons scenario A. On the other hand, in the case where the wavefunctions have different symmetry (on the right), bosons would prefer scenario B, while fermions scenario D. In this case both scenarios A and C have no preferences for bosons or fermions. These results are consistent with the generic principles of the previous section.

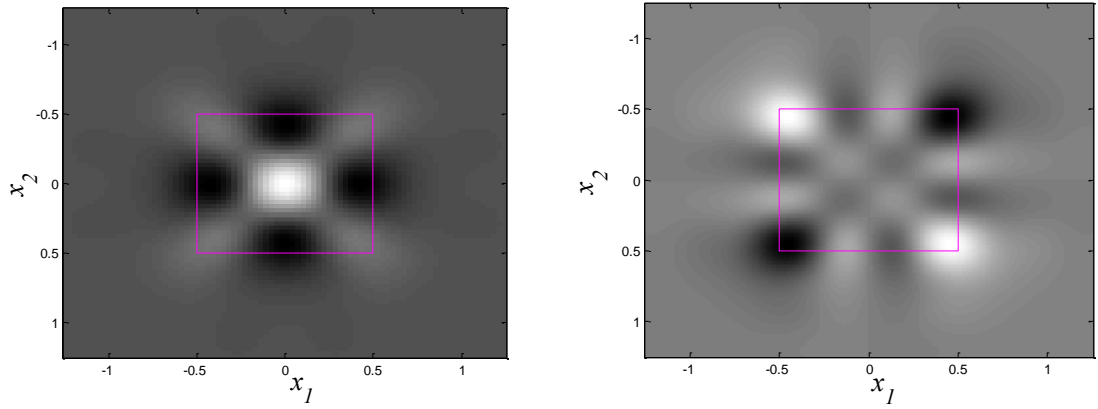

Figure 7: Gray levels presentation of $\delta(x_1, x_2; t = 0.03)$ for wavefunctions with the same symmetry (left) and different symmetry (right). White stands for $\Delta > 0$, i.e. more bosons, while Black stands for , $\Delta < 0$, i.e., more fermions.

## Summary and discussion

The scenario, in which a pair of particles is released from a quantum trap, was investigated for bosons and fermions. We show that basically there are four possible scenarios. Three of which correspond to escape probabilities, and each one of them have a unique temporal dynamics. The escape scenario can be described only by all three, focusing on a single scenario (as is done in many places) is a partial description of the dynamics. It was shown that the initial symmetry of the wavefunctions of the particles is crucial to the specific dynamics of the escape from the trap. The naive picture, in which bosons attract each other and fermions repel each other is a simplistic description, which cannot predict even *qualitatively* the behavior of the pair.

As one would expect, there is a higher chance that a Boson pair would remain in trap than a Fermion one. Similarly, there is a higher chance of finding a Boson pair escaping to the same direction (left or right). However, one of the main surprising conclusion of this work is, that if the two particles' wavefunctions have the same symmetry (either even or odd(12)), then there is a higher chance for a bosons pair (as opposed to fermions) to escape from the trap to *opposite directions*, as if they repel each other.

It was shown that the source of this effect comes directly from the particles statistics, and is totally generic. This effect was demonstrated in a numerical example, and an exact analytical expression for the short time expansion after the release was derived.

**Acknowledgment:** A.M. would like to thank Prof. Raizen and the Center for Nonlinear Dynamics at the University of Texas, Austin, for their hospitality, when this work was done.